\documentstyle[aps,multicol,float,epsfig,prb]{revtex}

\begin{document}

\title
{
Ferromagnetism and large negative magnetoresistance 
in Pb doped Bi-Sr-Co-O misfit-layer compound
}
\author{
I. Tsukada\footnote{Present address: Central Research Institute of 
Electric Power Industry, 2-11-1 Iwado-kita, Komae-shi, Tokyo 201-8511, JAPAN. 
E-mail: ichiro@criepi.denken.or.jp}, T. Yamamoto, M. Takagi, 
T. Tsubone, S. Konno, 
and K. Uchinokura\footnote{Also Department of Advanced Materials Science, 
The University of Tokyo.}
}
\address{
Department of Applied Physics, The University of Tokyo, 
7-3-1 Hongo, Bunkyo-ku, Tokyo 113-8656, JAPAN 
}
\date{\today}
\maketitle

\begin{abstract}
Ferromagnetism and accompanying large negative magnetoresistance in 
Pb-substituted Bi-Sr-Co-O misfit-layer compound are investigated 
in detail. 
Recent structural analysis of (Bi,Pb)${}_2$Sr${}_{3}$Co${}_2$O${}_9$, 
which has been believed to be a Co analogue of 
Bi${}_2$Sr${}_2$CaCu${}_2$O${}_{8+\delta}$, 
revealed that it has a more complex structure including a CoO${}_2$ 
hexagonal layer [T. Yamamoto {\it et al.}, Jpn. J. Appl. Phys. 
{\bf 39} (2000) L747]. 
Pb substitution for Bi not only introduces holes into the conducting 
CoO${}_2$ layers but also creates a certain amount of localized spins. 
Ferromagnetic transition appears at $T$ = 3.2~K with small spontaneous 
magnetization along the $c$ axis, and around the transition temperature 
large and anisotropic negative magnetoresistance was observed. 
This compound is the first example which shows ferromagnetic long-range 
order in a two-dimensional metallic hexagnonal CoO${}_2$ layer.
\end{abstract}

\begin{multicols}{2}
\narrowtext


\section{Introduction}

The discovery of high-temperature (high-$T_c$) superconductors has stimulated 
a considerable amount of researches on new conducting metal oxides.
\cite{Bednorz1} 
Crystal structures of high-$T_c$ cuprates can be 
regarded as an alternating stack of conducting CuO${}_2$ layers and 
insulating blocking layers. 
Many efforts to look for a new blocking layer have been still paid for 
a discovery of new cuprate superconductors. 
On the other hand, to find a new conducting layer other than the CuO${}_2$ 
layer is also an exciting challenge toward an exotic superconductor, 
and Sr${}_2$RuO${}_4$ is one of the good examples. 
\cite{Maeno1} 
Bi-$M$-Co-O ($M$ = Ca, Sr, Ba) system was first investigated 
by Tarascon {\it et al.,} according to such an idea.
\cite{Tarascon1} 
The chemical composition of this system was reported as 
Bi${}_2M_3$Co${}_2$O${}_9$, and the structure was believed to be 
as a Bi${}_2$Sr${}_2$CaCu${}_2$O${}_{8+\delta}$ (Bi2212) type one. 
The system becomes metallic by changing $M$ site from Ca to Ba.
\cite{Tarascon1,Watanabe1,Terasaki1} 
The chemical composition of Bi : Sr : Co : O = 2 : 3 : 2 : 9 
gives a valence of Co ions as 3+, and the magnetic susceptibility 
suggests that most of the Co${}^{3+}$ ions are in a low-spin state 
($S$ = 0) with fully occupied $t_{2g}$ levels and empty $e_g$ levels. 
Since the energy gap between these levels makes Bi${}_2M_3$Co${}_2$O${}_9$ 
a band insulator as LaCoO${}_3$,
\cite{Raccah1} 
the metallic conductivity induced by different alkaline-earth ions with 
the {\em same} valence cannot be accounted for straightforwardly. 
Another approach to introduce carriers by Pb substitution for Bi has been 
accomplished for both bulk and thin-film samples of Bi-Sr-Co-O system.
\cite{Tsukada1,Yamamoto1} 
However, the amount of the doped carriers did not correspond quantitatively 
to the amount of doped Pb, 
so that the mechanism of carrier doping has remained unclear for a long time. 

Recently we found that the structure of so-called 
(Bi,Pb)${}_2$Sr${}_3$Co${}_2$O${}_9$ phase is completely different from that of 
Bi-2212 superconductor. The detailed structural analysis has been done 
by x-ray and electron diffractions on (Bi,Pb)${}_2$Sr${}_3$Co${}_2$O${}_9$ 
single crystals with various Pb concentration, 
and this system was found to have misfit layer structure.
\cite{Yamamoto2} 
Our results are very similar to those reported by 
Leligny {\it et al.} for Pb-free sample,
\cite{Leligny1}
in which they stated that the chemical composition is approximately written 
as [Bi${}_{0.87}$SrO${}_2$]${}_2$[CoO${}_2$]${}_{1.82}$. 
Though the complete structure analysis on what we have called 
Bi${}_2$Sr${}_3$Co${}_2$O${}_9$ is not yet done, 
it is very probable that it has basically the same structure as that of 
[Bi${}_{0.87}$SrO${}_2$]${}_2$[CoO${}_2$]${}_{1.82}$,
\cite{Leligny1} 
where Pb is partially substituted for Bi.   
Now, all the experimental results should be reconsidered upon this structure. 
The most important modification from the old premise is that the Co ions 
in the conducting CoO${}_2$ layer form a triangular lattice 
instead of a rectangular lattice. 
In that sense, recently discovered large thermoelectric material 
NaCo${}_2$O${}_4$,
\cite{Terasaki2}
which has a hexagonal CoO${}_2$ layers, becomes a good reference to 
consider the transport and magnetic properties of (Bi,Pb)-Sr-Co-O system.

In this paper, we report transport and magnetic properties of a heavily 
Pb-doped single crystal that shows a ferromagnetic long-range order 
below 3.2~K. 
One of the characteristics of the ferromagnetic state is 
high anisotropy. 
Magnitude of the spontaneous magnetization is not so large as that 
observed in perovskite (La,Sr)CoO${}_3$. 
Our results rather suggests that the weak-ferromagnetic long-range order 
is the origin of the spontaneous magnetization, 
which may be closely related to the direction of each CoO${}_6$ octahedron 
in the CoO${}_2$ layer.

\begin{figure}
\includegraphics*{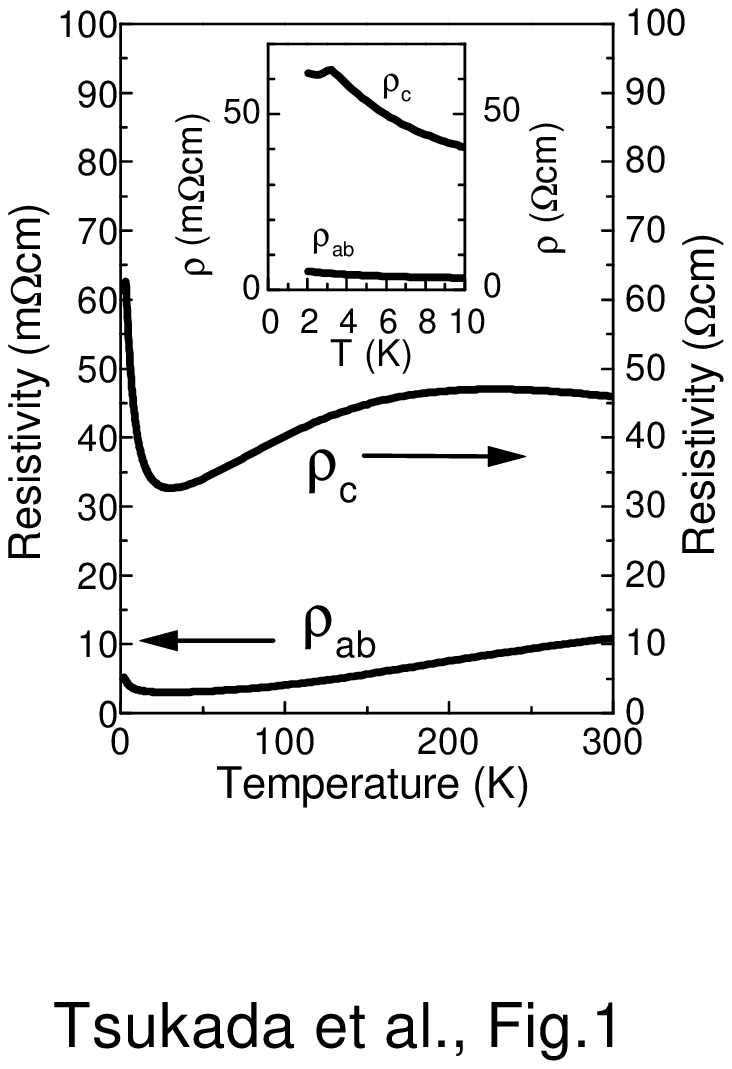}
\caption{
The temperature dependence of the in-plane and the out-of-plane 
resistivities. Note that the unit of the right ordinate for $\rho_c$ 
is three orders of magnitude larger than that of the left ordinate for 
$\rho_{ab}$. The inset shows the low-temperature behavior of $\rho_{ab}$ 
and $\rho_c$. 
}
\label{Fig.1}
\end{figure}

\section{Experimental Results}

\subsection{Sample Preparation}

The single-crystal sample was prepared by a floating-zone method 
using a polycrystalline rod with the nominal composition, 
Bi : Pb : Sr : Co = 1.2 : 0.8 : 2.0 : 2.0. 
Pb is easily evaporated during the growth process 
and the actual composition of the obtained single crystal 
determined by inductively coupled plasma-atomic emission 
spectroscopy is 
Bi : Pb : Sr : Co = 1.52 : 0.47 : 2.13 : 2.00. 
The ratio of (Bi+Pb) : Sr : Co = 1.99 : 2.13 : 2.00 
shows an excellent agreement with the reported 
chemical composition for a Pb-free sample as Bi : Sr : Co = 
1.91 : 2.19 : 2.00.
\cite{Leligny1} 
The crystal symmetry can be the same as that reported 
for the crystal with similar chemical composition:
\cite{Yamamoto2} 
C-centered monoclinic with $a$ = 4.92~{\AA}, $b_{RS}$ = 5.23~{\AA}, 
$b_{H}$ = 2.8~{\AA}, $c$ = 15.05~{\AA}, and $\beta$ = 93.55${}^{\circ}$, 
where the subscriptions $RS$ and $H$ denote (Bi${}_{1-x}$Pb${}_x$Sr)O${}_2$ 
rock-salt layer and CoO${}_2$ hexagonal layer, respectively.
\cite{c-axis} 
Transport and magnetic  properties were measured by commercial physical 
properties measurement system (PPMS, Quantum Design) and SQUID magnetometer 
($\chi$-MAG, Conductus).

\subsection{Resistivity}

The resistivity measurements were carried out along both the 
in-plane and the out-of-plane directions as shown in Fig.~\ref{Fig.1}. 
The metallic conductivity was observed along the in-plane ($\rho_{ab}$) 
direction. 
The resistivity at room temperature was around 10~m${\Omega}$cm, 
and d${\rho} / $d$T >$ 0 behavior was observed down to 25~K. 
The magnitude of the resistivity is quite high in comparison with a 
clean conventional metal, 
which was also observed in the thin-film samples.
\cite{Yamamoto1} 
The preliminary heat-capacity measurement gives a large electronic 
specific-heat coefficient as $\gamma$ = 98.2~mJ/Co-mol K${}^2$, 
which indicates that the effective mass of carriers are strongly enhanced 
implying the presence of strong electron correlation in this compound. 
The detailed results will be reported elsewhere.
\cite{Yamamoto3}

The out-of-plane resistivity (${\rho}_c$) shows rather complex behavior 
(Fig.~\ref{Fig.1}). 
At room temperature, ${\rho}_c$ is larger than ${\rho}_{ab}$ by four 
orders of magnitude. 
There is a broad peak around 200~K and then decreases down to 30~K. 
Such a broad-peak feature is similar to that observed in 
Sr${}_2$RuO${}_4$
\cite{Lichtenberg1} 
and NaCo${}_2$O${}_4$,
\cite{Terasaki2} 
but is different from that observed 
in Bi-2212 high-$T_c$ superconductor.
\cite{Ito1} 
In-plane metallic and out-of-plane insulating behaviors, which is believed 
to be strong indication of {\em charge confinement\/} characteristic 
to many high-$T_c$ cuprates, were not found, and thus the system is 
better described as a highly anisotropic 3D metal. 
The broad peak of ${\rho}_c$ may be determined by the competition between 
the mean-free-path along the $c$-axis electron transport and the separation 
of the neighboring conducting CoO${}_2$ planes. 
There is an upturn below 20~K in both ${\rho}_{ab}$ and 
${\rho}_c$. The inset of Fig.~\ref{Fig.1} shows a closer look at the 
low-temperature resistivity. 
A smooth increase in ${\rho}_{ab}$ toward 2~K shows a good contrast to 
the cusp in ${\rho}_c$ at 3.2~K. 
This temperature corresponds to the ferromagnetic transition as will be 
clarified by the magnetization measurement.

\subsection{Magnetization}

\begin{figure}[t]
\includegraphics*{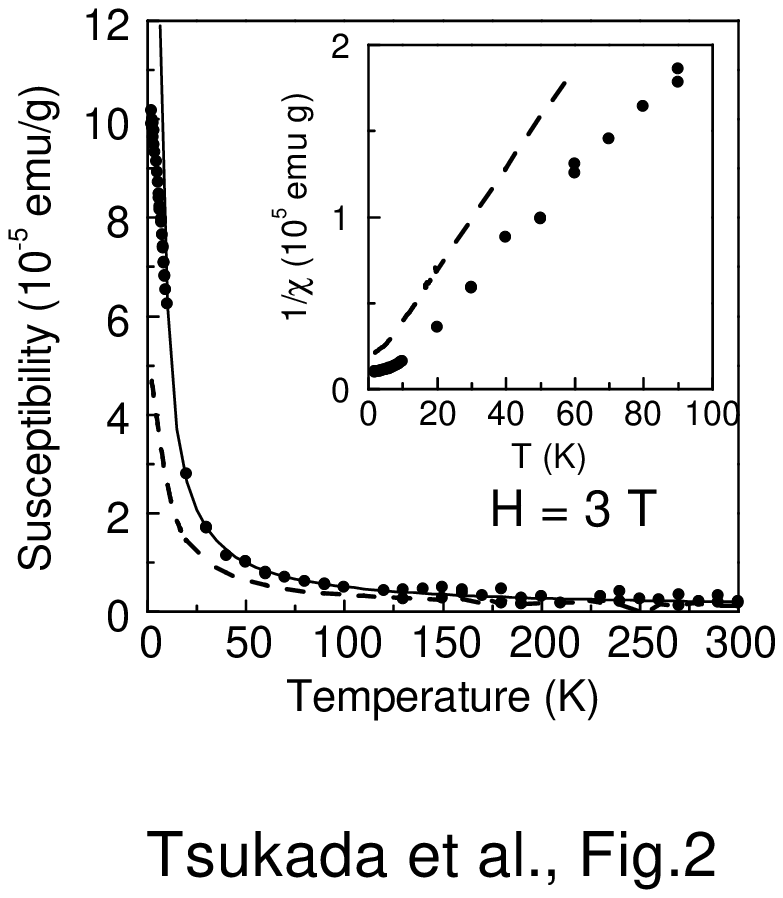}
\caption{
Temperature dependence of the susceptibility with $H$ = 3~T 
along the $c$ axis. The solid line is the Curie-Weiss fit 
with constant term to the data. The dashed line shows the data 
of the Pb-free sample taken from Ref. 7). 
The inset shows the temperature dependence of 
the inverse susceptibility.}
\label{Fig.2}
\end{figure}

The temperature dependence of susceptibility is shown in Fig.~\ref{Fig.2} 
with that of Pb-free samples. 
The magnetic field is applied perpendicular to the $ab$ plane. 
Because of the small sample mass, the data above 100~K becomes noisy 
even under relatively high-field, $H$ = 3~T. 
There is no trace of magnetic long-range order in this temperature region. 
The inset of Fig.~\ref{Fig.2} shows the inverse susceptibility. 
1/$\chi$ roughly shows $T$-linear behavior above 20~K. 
It is easily seen that the Pb substitution increases the magnitude 
of effective Bohr magneton. 
We fit the data by a sum of the Curie-Weiss term and the 
temperature-independent constant term (solid lines). 
The effective number of Bohr magneton per Co ion was estimated as 
$P_{eff}$ = $\sqrt{4S(S+1)}$ = 1.28 
which leads to a small spin number consistent with that most of the 
Co ions are in a low-spin state. 
The extrapolation from the high-temperature $T$-linear part 
to $T$ = 0~K meets the horizontal axis approximately at $T$ = 4~K, 
suggesting the ferromagnetic order around this temperature, 
which is different from the Pb-free sample.

\begin{figure}[t]
\includegraphics*{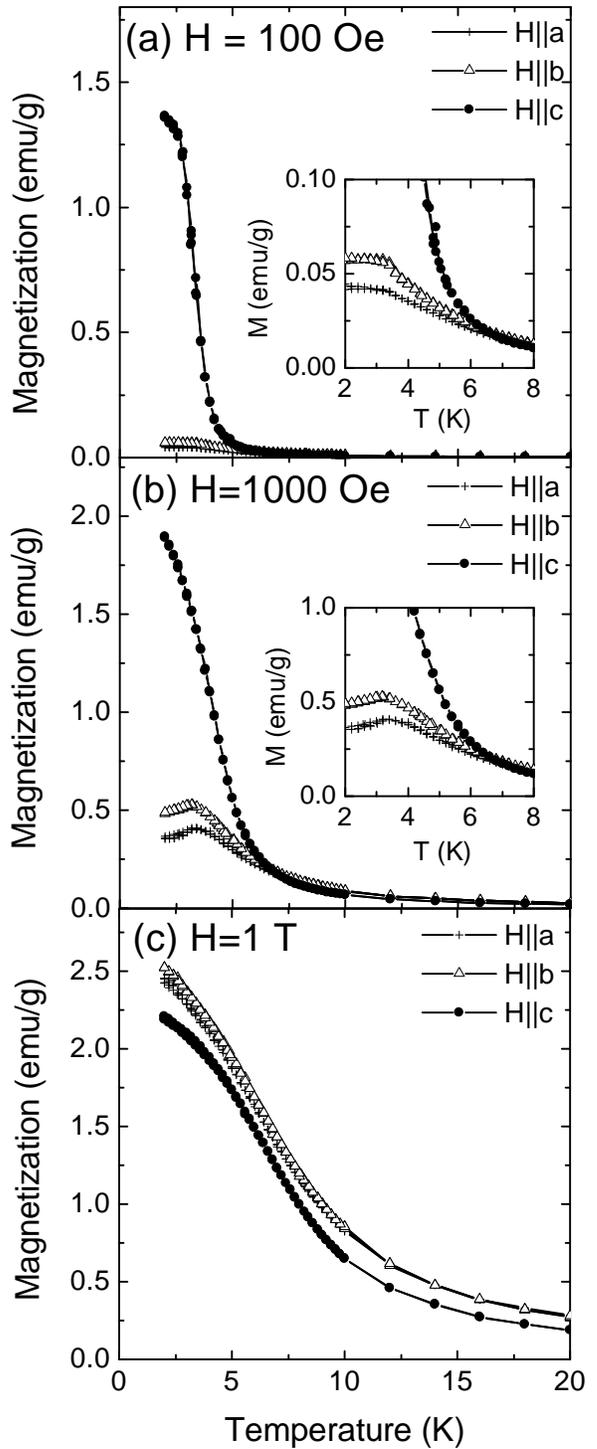}
\caption{
Temperature dependence of the magnetization along the 
three principal axes at (a) 100~Oe, (b) 1000~Oe, and (c) 1~T. 
At 100~Oe and 1000~Oe, the magnetization is highly anisotropic, 
while at 1~T, it becomes almost isotropic.}
\label{Fig.3}
\end{figure}

Under low magnetic field, the spontaneous magnetization actually appears 
along the $c$ axis below 3.2~K, 
where we observed the cusp in $\rho_c$. 
Figure~\ref{Fig.3} shows the magnetization along the three principal axes 
measured with three different fields. 
In Fig.~\ref{Fig.3}(a), the evolution of the spontaneous 
magnetization is observed along the $c$ axis. 
For $H {\parallel} a$ and $H {\parallel} b$, magnetizations are saturated 
below $T_c$. 
Saturated value is far smaller than that along the $c$ axis. 
Similar anisotropy remains even at 1000~Oe, but disappears at 1~T. 
At 1~T, magnetization becomes almost isotropic along the three principal 
axes. The slight difference between the magnetizations along the in-plane 
and the out-of-plane directions is probably due to the difference 
in $g$ factors.

\subsection{Magnetoresistance}

\begin{figure}[t]
\includegraphics*{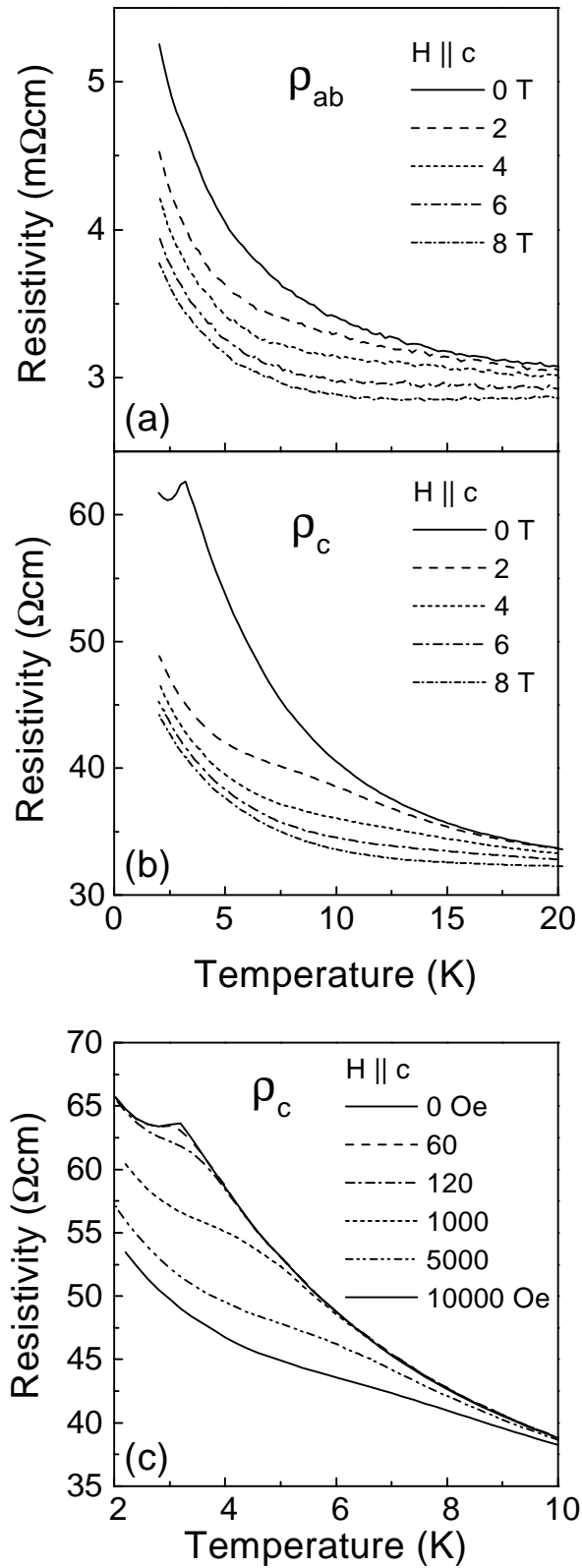}
\caption{
Low-temperature magnetoresistance along both (a) the in-plane and 
(b) the out-of-plane directions. The low-field behavior of $\rho_c$ is 
shown in (c).}
\label{Fig.4}
\end{figure}

The presence of a cusp in $\rho_c$ at $T_c$ indicates a strong coupling 
between the conduction carriers and the background magnetic moment. 
We measured the in-plane and the out-of-plane resistivities under several 
magnetic fields applied along the $c$ axis as shown in Fig.~\ref{Fig.4}, 
in which large negative magnetoresistance was observed. 
At 8~T, $\rho_{ab}$ decreases to 71{\%} of the value at 0~T. 
The change is as large as that observed in ferromagnetic (La,Sr)CoO${}_3$.
\cite{Yamaguchi1} 
However, $\rho_{ab}$ does not show a distinct change in its slope 
at $T_c$, and the reduction of the resistivity below $T_c$ seems to be 
a parallel shift. 
This is qualitatively different from that of (La,Sr)CoO${}_3$ or 
the colossal magnetoresistance (CMR) manganites.

The negative magnetoresistance along the out-of-plane direction is 
similar to that of CMR materials. 
As shown in Fig.~\ref{Fig.4}(b), $\rho_c$ shows more drastic
change with the magnetic field. 
Initial resistivity suppression up to 2~T is remarkable (Fig.~\ref{Fig.4}(b)), 
and $\rho_c$ becomes less than 80{\%} of the initial value at $T$ = 2~K. 
Above 2~T, the reduction of the resistivity becomes more gradual, 
and at 8~T $\rho_c$ is suppressed to less than 72{\%} of the initial value. 
This anisotropy is probably related to the anisotropic magnetization. 
One of the remarkable features is that the steep increase of the $c$-axis 
magnetization {\em does not} induce large negative magnetoresistance 
in both $\rho_{ab}$ and $\rho_c$. 
For example, $\rho_c$ remains almost unchanged with the magnetic 
field up to 120~Oe as shown in Fig.~\ref{Fig.4}(c), 
except for a narrow region near $T_c$. 
After the field exceeds 120~Oe, the resistivity begins to decrease 
significantly with increasing magnetic field.

The field-direction dependence provides us the details of the anisotropic 
nature. 
We measure the in-plane resistance under the magnetic field both 
parallel ($H {\parallel} ab$, $I {\perp} H$) and perpendicular 
($H {\perp} ab$, $I {\perp} H$) to the plane direction. 
In Fig.~\ref{Fig.5}, the magnetization and the in-plane resistance ($R_{ab}$) 
are plotted as functions of a magnetic field both parallel 
and perpendicular to the plane.
In Fig.~\ref{Fig.5}(a) the $c$-axis magnetization initially shows a 
steep increase to 0.06~$\mu_B$/Co site, and then turns to show a gradual 
increase. In contrast to it, $R_{ab}$ does not change drastically with 
the initial evolution of the magnetization, and it significantly decreases 
with the following moderate increase of the $c$-axis magnetization. 
Figures~\ref{Fig.5}(c) and \ref{Fig.5}(d) show a closer look at the 
magnetization and $R_{ab}$ at $H {\parallel} c$ below 1~kOe. 
$R_{ab}$ roughly keeps its initial value during the steep 
increase of magnetization, and then it begins to decrease 
after the magnetization changes its field dependence 
above 120~Oe (indicated by a dotted line).

Similar plateau in $R_{ab}$ is also observed when we apply the field 
along the in-plane direction (Figs.~\ref{Fig.5}(e) and \ref{Fig.5}(f)). 
For $H$ ${\parallel}$ plane, we did not observe negative magnetoresistance 
near $H$ = 0~T. 
On the contrary, small positive magnetoresistance was observed below 2~kOe. 
The corresponding magnetization evolution with $H$ ${\parallel}$ plane is 
rather smooth, and it is difficult to assign where the magnetization 
shows a kink. However, the initial evolution of the 
magnetization seems to be proportional to the applied field (indicated 
by a dashed line in Fig.~\ref{Fig.5}(e)), and starts to deviate from the 
linear relation roughly above 3~kOe. 
At the same field, $R_{ab}$ begins to show large negative magnetoresistance, 
and we think that the low-field and high-field behaviors are qualitatively 
different as was observed in $H$ $\perp$ plane configuration.

To summarize our observation, 
1) the low-field steep increase of magnetization does not induce negative 
magnetoresistance, 
and 2) the additional magnetization following the low-field one induces 
negative magnetoresistance. 
This result implies that two dominant contributions to 
the magnetization should be considered.

\section{Discussion}

\subsection{Pb substitution}

First we discuss the origin of magnetic moment induced by 
Pb substitution for Bi. For this purpose, 
we assume [Bi${}_{0.87}$SrO${}_2$]${}_2$[CoO${}_2$]${}_{1.82}$ 
as the chemical composition of the parent compound.
\cite{Leligny1} 
Since Bi${}^{3+}$ and Sr${}^{2+}$ ions are nonmagnetic, 
most of the magnetic properties are attributed to Co ions. 
The reported chemical composition gives average valence of Co ion as +3.33. 
Two scenarios then become possible to account for this non-integer valence: 
1) one of three Co ions takes localized 4+ state while the others take 
localized 3+ state, 
or 2) one mobile hole is doped per three Co${}^{3+}$ sites. 
In the previous magnetization measurement by Tarascon {\it et al.}, 
it was suggested that most of the Co ions are in the 3+ state and 
takes low-spin state with $S$ = 0.
\cite{Tarascon1} 
Even though their treatment of the 
chemical composition is possibly incorrect, absence of the large 
magnetic moment is still an experimental fact. 
Thus the former case does not seem to be the current case. 
The latter is more probable, and we can imagine that holes have been 
already doped in a Pb-free sample. 
Note that metallic conductivity was observed even in the Pb-free samples
\cite{Tsukada2}, 
which suggests the presence of holes even in the parent compound. 
Recent photoemission experiment also supports the same conclusion.
\cite{Mizokawa1}

It is interesting to see that the magnetic susceptibility of NaCo${}_2$O${}_4$, 
which has a similar CoO${}_2$ hexagonal layer, is quite similar to our data. 
The nominal composition of NaCo${}_2$O${}_4$ gives averaged Co 
valence as 3.5+. If we adopt localized-moment picture, half of the 
Co ions should be in the 4+ state that will have $S$ = 1/2 or $S$ = 5/2 
for low-spin or high-spin configurations, respectively. 
However, the magnetic susceptibility is far smaller than that expected 
for such localized picture.
\cite{Tanaka1}
Recent local density approximation (LDA) calculation for NaCo${}_2$O${}_4$ 
implies that Co $t_{2g}$ band is located at the Fermi level ($E_F$), 
and dominates all the magnetic properties of NaCo${}_2$O${}_4$.
\cite{Singh1} 
If we apply this expectation also to (Bi,Pb)-Sr-Co-O because of the 
similarity of the crystal structure of hexagonal CoO${}_2$ plane, 
the doped holes are located at Co $t_{2g}$ band, and 
we do not have to take the contribution of other bands into account.

Then how the Pb substitution changes the spin system? 
One of the important roles of the Pb substitution is to introduce 
additional holes. 
Using the measured chemical composition and assuming that the oxygen content 
is the same as that in the parent compound, we can estimate the average 
valence of Co ions as 3.52+ for our sample. 
The number of holes fairly increases and the electronic conduction is 
enhanced, which has been already observed in our previous studies.
\cite{Tsukada1,Yamamoto1}

Another significant role of the Pb substitution is 
to modify the crystal parameters. 
As was reported, Pb substitution induces a discontinuous decrease in the 
$b$-axis length of the Bi${}_{0.87}$SrO${}_2$ layer and expands the $c$-axis 
length simultaneously.
\cite{Yamamoto2}
Since we have no data on the structural change in the CoO${}_2$ layer, 
at present it is difficult to discuss the effect of the structural change 
to the electronic state. 
As is shown in Fig.~\ref{Fig.2} and ref. 7, however, the increase 
of the effective Bohr magneton above $T_c$ by the Pb substitution suggests 
that the Pb substitution also increases the number of localized spins, and 
such spins may contribute to the formation of ferromagnetic long-range order.

Before discussing the relation between ferromagnetic 
long-range order and magnetoresistance in the next section, 
it is better to mention the possibility of itinerant ferromagnetism 
in this system. 
It is well known that ferromagnetic long-range order can also be observed 
in an itinerant electron system. In our case, the $c$-axis magnetization 
seems to be explained also by weak ferromagnetism of itinerant 
electrons. 
However, the anisotropic magnetization in the weak-ferromagnetic state 
is hardly explained by the itinerant ferromagnetism. 
To our knowledge, there has been no theory of itinerant ferromagnetism 
with highly anisotropic magnetization. Although this does not exclude 
the possibility that our highly two-dimensional system exhibits an 
itinerant ferromagnetism by unknown mechanism, it is better to explain 
our results based on the localized-spin model.

\subsection{Ferromagnetic long-range order and magnetoresistance}

Once we assume that the localized moments are induced by Pb substitution, 
we can explain the field dependence of the magnetization. 
The magnitude of the spontaneous magnetization is quite small below $T_c$, 
and subsequently the field-induced magnetization keeps increasing at least 
up to 7~T. 
This field dependence strongly indicates that the system goes into 
a weak-ferromagnetic state. 
In a conventional weak ferromagnet a certain antiferromagnetic unit 
has a finite ferromagnetic moment, and a magnetization process is separated 
in two parts. 
In a low-field region, inversion of the direction of the weak ferromagnetic 
moment gives steep evolution of the magnetization, while at high-field region 
each spin tends to be aligned along the magnetic field. 
At the present case, spins are probably confined roughly within the $ab$ 
plane, and are tilted a little toward the $c$-axis direction. 
Then a single CoO${}_2$ layer can have a weak ferromagnetic moment 
along the $c$ direction. 
The absence of net bulk magnetization at $H$ = 0~Oe indicates that 
such weak ferromagnetic layers are stacked randomly or alternatingly 
along the $c$ axis. 
When the $c$-axis field reaches 120~Oe, these weak ferromagnetic moments 
point to the same direction. 
Above the critical field each spins begins to be aligned along the $c$ axis, 
which corresponds to an increase in the magnitude of 
weak-ferromagnetic moment. 

\end{multicols}

\widetext

\begin{figure}
\includegraphics*{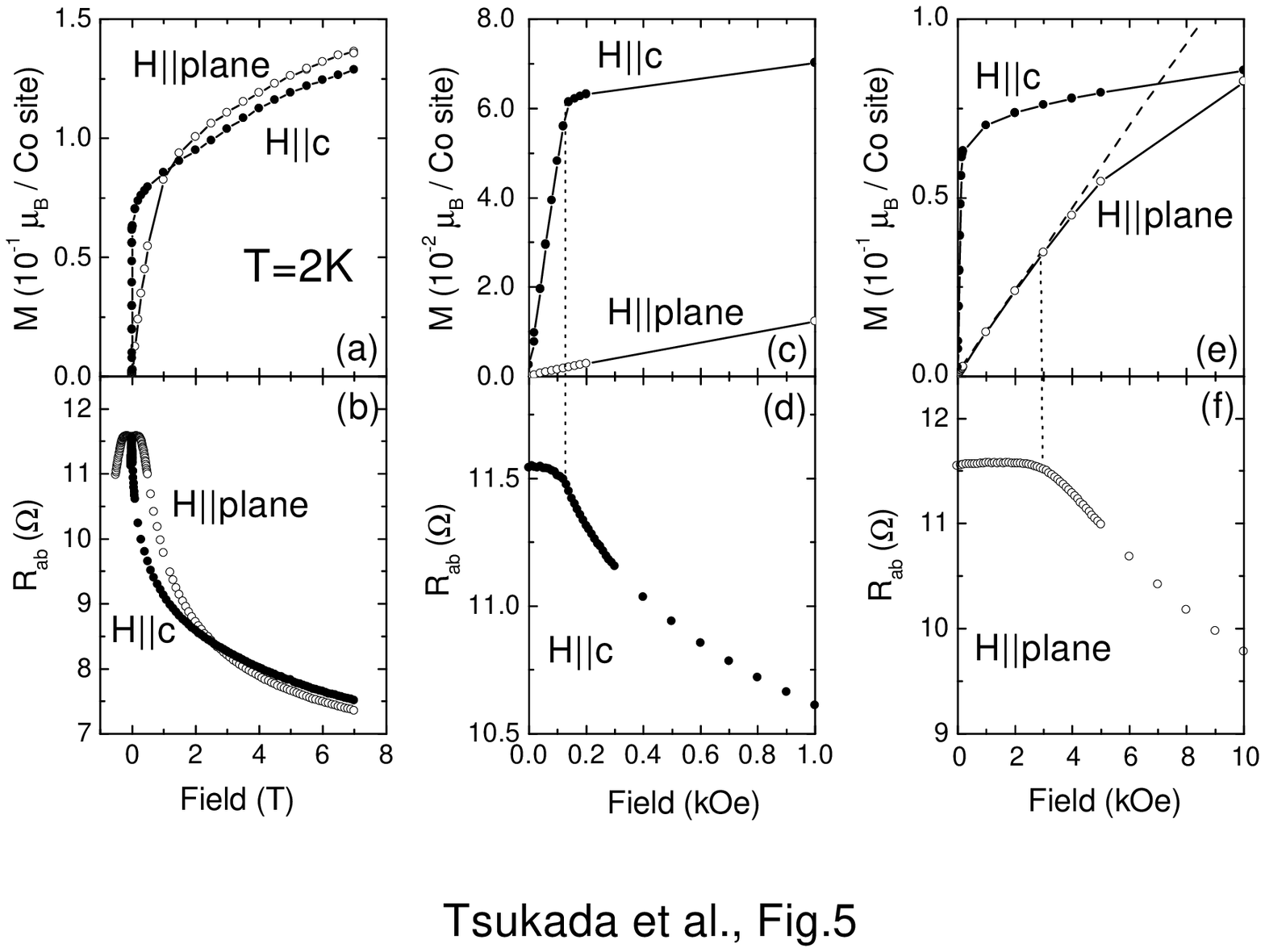}
\caption{
Magnetic field dependence of the magnetization and $R_{ab}$  
along both for $H {\parallel}$ plane and $H {\parallel} c$. 
(a) and (b) show the data up to 7~T. (c) and (d) show a closer look 
around the saturation field under $H || c$ where the magnetization 
and $R_{ab}$ change their field 
dependence. 
(e) and (f) show a closer look around the saturation field under 
$H {\parallel}$ plane where $R_{ab}$ changes its field dependence.}
\label{Fig.5}
\end{figure}

\begin{multicols}{2}
\narrowtext

The magnetization process along the plane direction is rather moderate 
as shown in Fig.~\ref{Fig.5}(e). However, a careful look at the 
magnetization allows us to divide its behavior in two regions. 
The linear evolution of the magnetization is observed up to 3~kOe, 
where the in-plane resistance begins to decrease. 
Above 3~kOe, magnetization deviates from this linear relation (dashed line). 
This feature is different from that of a typical weak ferromagnet 
with strong easy-axis anisotropy. 
The presence of finite linear M/H regions along both directions during the 
field inversion process suggests a more complicated situation of Pb-doped 
Bi-Sr-Co-O, and  we expect that the hexagonal network of Co ions plays 
an important role.

Next we discuss the mechanism of large negative magnetoresistance. 
One of the characteristics of the magnetoresistive behavior is 
the resistance plateau below the critical field. 
The weak ferromagnetic state can explain this plateau. 
Since the low-field magnetization perpendicular to the 
$ab$ plane is considered to be an ordering process for the weak 
ferromagnetic moment of each CoO${}_2$ layer. 
The change in the direction of weak ferromagnetic moment is probably 
accomplished by the inversion of each spin. 
This process {\em does not} modify the relative angle of 
spins in the same CoO${}_2$ layer, and therefore, the in-plane resistance 
does not need to be influenced. 
A similar process has been discussed in a layered perovskite manganite 
(La,Sr)${}_3$Mn${}_2$O${}_7$.
\cite{Kimura1} 
In such a manganite, a ferromagnetic moment is attributed to each layer, 
and there is a transition at which the long-range order 
of this moment develops along the $c$ directions. At this transition 
temperature, the in-plane resistivity does not change much while 
the out-of-plane resistivity drastically changes. 
Once the field exceeds the critical value, each spin begins to rotate 
toward the same direction, and as a result, in-plane resistance can decrease 
under a strong coupling between the itinerant holes and the localized 
moments. 
When the field is applied parallel to the plane, large magnetoresistance 
begins to appear above 3~kOe. This indicates that spins do not drastically 
change their relative angle below 3~kOe. 
As we mentioned before, the in-plane magnetization process is not simple, 
but again the hexagonal lattice may play a key role. 
We need further investigations of this system to reveal the 
magnetic structure with various experimental techniques, 
such as NMR, ESR, and neutron scattering.

The origin of the weak ferromagnetism is still an open question. 
In a conventional weak ferromagnetic state, spin-spin interaction is 
basically antiferromagnetic, and the secondary interaction tend to 
cant spins toward a certain direction. Two kinds of the origin of 
such spin canting have been known: antisymmetric exchange interaction 
and single-ion anisotropy. The former was first discussed by 
Dzyaloshinskii,
\cite{Dzyaloshinskii1} 
and later microscopically formulated by Moriya.
\cite{Moriya1} 
However, this type of antisymmetric interaction can exist only when 
the crystal symmetry of two neighboring magnetic ions is sufficiently low. 
Since we do not consider that all the Co ions have localized magnetic moments, 
it is difficult to imagine that the antisymmetric interaction plays an 
important role between such diluted magnetic moments. 
The latter single-ion anisotropy is also inapplicable if the induced 
moment is attributed to a low-spin Co${}^{4+}$ ion. 
The orbital angular momentum of the low-spin Co${}^{4+}$ state is probably 
quenched, and thus the single-ion anisotropy cannot exist. 
Only when Co${}^{4+}$ ion is in a high-spin state, this scenario is possible. 
Preliminary photoemission spectroscopy suggests that the Co${}^{4+}$ ion 
induced by Pb-substitution is in the low-spin state,
\cite{Mizokawa1} and thus we cannot simply apply this mechanism.

\section{Conclusion}

To summarize we have investigated the transport and the magnetic 
properties of single-crystal 
(Bi${}_{1.52}$Pb${}_{0.47}$)Sr${}_{2.13}$Co${}_{2.00}$O${}_y$. 
The large anisotropy is found in the electronic conduction 
parallel and perpendicular to the plane, which reaches to 10${}^4$. 
The ferromagnetic transition takes place at very low temperature 
with small spontaneous magnetization. 
The large negative magnetoresistance is observed around the transition 
temperature, which strongly depends on the direction of magnetic field. 
Recently several layered compounds including hexagonal CoO${}_2$ layers 
have been investigated from the viewpoint of electric and magnetic properties: 
such as NaCo${}_2$O${}_4$ (ref. ~11), 
and Ca${}_3$Co${}_4$O${}_9$ misfit layer compound.
\cite{Masset1} 
The similarity between Ca${}_3$Co${}_4$O${}_9$ and our system is interesting, 
and from these compounds the common physics that is characteristic 
to the hexagonal metallic CoO${}_2$ layer will be deduced

\section*{Acknowledgment}

We thank I. Terasaki, Y. Tokura, T. Mizokawa and J. M. Tranquada for 
enlightening discussions. 
This work was supported in part by the Grant-in-Aid for Scientific Research on 
Priority Area ``Mott transition,'' Grant-in-Aid for COE Research 
``SCP coupled system,'' Grant-in-Aid for Encouragement of Young Scientists 
(I. T.) from the Ministry of Education, Science, Sports and Culture.


\end{multicols}


\begin{references}
\bibitem{c-axis} Exactly saying, the $c$-axis direction is not perpendicular 
to the $ab$ plane because of the monoclinic crystal symmetry. However, we use 
the $c$-axis direction to denote the direction perpendicular to the $ab$ plane 
throughout the paper, because it makes comparison with the previous 
literatures easy.
\end{references}
\end{document}